\documentstyle[aps,epsf,prl,multicol]{revtex}
\begin{document}
\title{Resistance of a 1D random chain: \ \\
Hamiltonian version of the transfer matrix approach}

\author{V.Dossetti-Romero and F.M.Izrailev}
\address{Instituto de F\'{\i}sica, B.U.A.P., Apdo. Postal J-48,
72570 Puebla, M\'exico}
\author{A.A.Krokhin}
\address{Instituto de F\'{\i}sica, B.U.A.P., Apdo. Postal J-48,
72570 Puebla, M\'exico \ \\
Department of Physics, University of North Texas,
P.O. Box 311427, Denton, TX 76203}

\date{\today}
\maketitle

\begin{abstract}
We study some mesoscopic properties of electron transport by employing
one-dimensional chains and Anderson tight-binding model. Principal attention
is paid to the resistance
of finite-length chains with disordered white-noise potential. We develop
a new version of
the transfer matrix approach based on the equivalency of a discrete
Schr\"odinger equation and a two-dimensional Hamiltonian
map describing a parametric kicked oscillator. In the two limiting cases of
ballistic and localized regime we demonstrate how analytical results for
the mean resistance and its second moment can be derived directly from
the averaging over classical trajectories of the Hamiltonian map. We
also discuss the implication of the single-parameter
scaling hypothesis to the resistance.
\end{abstract}

\begin{multicols}{2}


\section{Introduction.}

In recent years, the study of one-dimensional (1D) tight-binding models
with diagonal disorder
has led to new important results. The growing interest in the 1D
disordered models is mainly due to two reasons.
First, it was found that specific correlations in (random) potentials
give rise to delocalized states. Early demonstrations of this effect are
related to the so-called dimer model \cite{dimer} that is specified by
{\it short-range}
correlations. It was shown that for particular
{\it discrete} values of the electron energy
the localization length diverges, in contrast to a common belief that any
randomness in the
potential leads in 1D geometry to an exponential localization of all
eigenstates. Recently, delocalized states have been experimentally
observed in random dimer superlattice \cite{Bellani}. Similar effects of
delocalization due to correlated disorder have been also predicted for
some special 1D models, see \cite{DR79} and references in \cite{IKU01}.

Further study of 1D random models with correlated potentials
produced even more exciting results. It was shown \cite{lyra,IK99}
that specific {\it long-range} correlations may result in an
appearance of the mobility edges. This means that a {\it
continuous} range of energy arises on one side of the mobility
edge, where the eigenstates are extended. In Ref.\cite{IK99} the
method for constructing such correlated random potentials has been
proposed. Using this method, one can relatively easily construct
specific random potentials that result in energy bands of a
complete transparency for finite samples.  The position and the
width of such windows of transparency can be controlled by the
form of the binary correlator of a weak random potential.

The role of long-range correlations has been studied in details
for the tight-binding Anderson-type model \cite{IK99}, and for the
Kronig-Penney model with randomly distributed amplitudes
\cite{KI00} and positions of delta-peaks \cite{IKU01}. These results
have been also extended to a single-mode waveguide with random
surface profiles \cite{IM01}. The predictions of the theory
\cite{IK99} have been verified experimentally \cite{KIKS00},
when studying transport properties of a single-mode
electromagnetic waveguide with point-like scatterers. The latter
were intentionally inserted into the waveguide in a way to
provide a random potential with slowly decaying binary correlator.
Very recently \cite{IM02} the existence of mobility edges was
predicted for waveguides with a finite number of propagating
channels (quasi-1D system) with long-range correlations in a random
surface scattering potential.

Another reason for the revival of the interest in 1D disordered models
is due to the revision of
the famous {\it single parameter scaling} hypothesis (SPS) \cite{AALR79} for
transport characteristics of disordered conductors. As is shown
in \cite{alt}, the assumption of a random character of fluctuations of
the {\it finite-length} Lyapunov exponent, originally
used \cite{ATAF80} in order to justify the SPS, turns out to be incorrect.
Specifically, it
was found that in the vicinity of the band edges the SPS is violated
\cite{alt}. This result is important both from the theoretical and experimental
viewpoint (see discussion
and references in \cite{alt}).

One of the effective tools to analyze 1D tight-binding models is the one
based on the Hamiltonian map (HM) approach that has been developed in
Refs.\cite{Hmap,KTI97,IRT98}. The key point of this approach is a transformation
that reduces a discrete 1D Schr\"odinger equation to the classical
two-dimensional Hamiltonian map. The properties of trajectories of
this map are related to transport properties of a quantum model.
The geometrical aspects of the HM approach turn out to be helpful
in qualitative analysis, as well as in deriving analytical
formulas.

With this method many known results can be easily
obtained for the localization length $l_\infty$.
In particular, a rigorous derivation of $l_\infty$ was made
in a general form for energy close to the energy bands of the
standard Anderson model with uncorrelated diagonal disorder  \cite{IRT98}.
This method has been also used in \cite{IK99} to derive a general
expression for the localization length for {\it any} weak potential.
All these studies refer to {\it infinite} samples and are based on
the evaluation of the {\it infinite-length} Lyapunov exponent that is determined by
the evolution of the classical Hamiltonian two-dimensional map.

In this paper we present further developments of the HM approach,
that allows us to obtain important transport characteristics in 1D random
models. The main attention is paid to the resistance of finite
samples that are described by the standard tight-binding Anderson
model with weak white-noise potential. We derive analytical
expressions for the mean value of the resistance and its second moment. These are
the quantities that are easy to measure experimentally. We also
discuss the implication of the SPS hypothesis for the resistance in
strongly localized regime.

\section{Hamiltonian map}

The stationary discrete Schr\"{o}dinger
equation for 1D tight-binding models with diagonal disorder can be written
as follows,
\begin{equation}
\psi_{n+1}+\psi_{n-1}=\left( E+\epsilon_{n} \right)\psi_{n},
\label{eq:1}
\end{equation}
where $E$ is the energy of an eigenstate and $\epsilon_{n}$ is the site potential.
It is known that this equation is equivalent to diagonalization of a tridiagonal
matrix \cite{SolSt} or to a classical Hamiltonian map for canonical variables
$x_{n}=\psi_{n}$ and $p_{n}=(\psi_{n}\cos\mu-\psi_{n-1})/\sin\mu$
(see, e.g.\cite{Hmap}),
\begin{equation}
\left( \begin{array}{c}
             x_{n+1} \\ p_{n+1}
        \end{array} \right) =
\left( \begin{array}{clcr}
             (\cos \mu + A_{n} \sin \mu) \,\,\,& \sin \mu \\
             (A_{n} \cos \mu - \sin \mu)  \,\,\, & \cos \mu
        \end{array} \right)
\left( \begin{array}{c}
             x_{n} \\ p_{n}
        \end{array} \right)
\label{eq:2}
\end{equation}
The canonical variables $x_{n}$ and $p_{n}$ can be considered as
position and momentum of a linear oscillator subjected to linear
periodic delta kicks. The amplitude of the $n$th kick depends on
the electron energy at the $n$th site potential,
$A_{n}=-\epsilon_{n}/\sin\mu$. In this approach, the amplitude
$\psi_{n}$ of an eigenstate at the $n$th site is given by the
position of the oscillator at time $t_{n} = n$. Therefore, global
properties of the eigenstates can be studied by exploring the time
dependence of the classical map (\ref{eq:2}). The energy $E$ is
related to the angle $\mu$ via $E=2\cos\mu$.

One can see that the representation (\ref{eq:2}) is, in essence, the Hamiltonian version
of the standard transfer
matrix method. Indeed, starting from two initial values $\psi_{0}$ and $\psi_{-1}$ one
can compute
$\psi_{n}$ and $\psi_{n-1}$ according to the recursion given by Eq.(\ref{eq:1}) or
Eq.\ (\ref{eq:2}).
Such a representation
allows one to determine the Lyapunov exponent $\Lambda$ by running
the trajectory of the map Eq. (\ref{eq:2}) from $n = 0$ to $n\rightarrow\infty$.
According to the standard definition (see, e.g. \cite{LGP88}),
the inverse of the Lyapunov exponent $\Lambda^{-1}(E)$ gives the localization length
$l_{\infty}(E)$ of the
eigenstate.

In many aspects the
Hamiltonian representation (\ref{eq:2}) is more convenient than the standard one based
on the original equation (\ref{eq:1}).
The effectiveness of the treatment of the map (\ref{eq:2}), instead of
(\ref{eq:1}), is clearly manifested if the action-angle
variables $\left(r,\theta\right)$ are introduced according to the following
transformation, $x=r\sin\theta$ and
$p=r\cos\theta$. Then the map Eq. (\ref{eq:2}) takes the following form,
\begin{equation}
\begin{array}{l}
\sin\theta_{n+1} = D_{n}^{-1} \left[ \sin\left(\theta_{n}-\mu\right)
- A_{n}\sin\theta_{n}\sin\mu \right], \\ \\
\cos\theta_{n+1} = D_{n}^{-1} \left[ \cos\left(\theta_{n}-\mu\right)
+ A_{n}\sin\theta_{n}\cos\mu \right] \nonumber
\end{array}
\label{eq:3}
\end{equation}
where
\begin{equation}
D_{n} = \frac{r_{n+1}}{r_n} =
\sqrt{1 + A_{n}\sin\left(2\theta_{n}\right) + A_{n}^{2}\sin^{2}\theta_{n}} .
\label{eq:4}
\end{equation}

Using action-angle variables (\ref{eq:3}), (\ref{eq:4}), the inverse localization
length is written in a quite simple form (see details in \cite{IRT98}),
\begin{eqnarray}
l_{\infty}^{-1} \equiv \Lambda  =  \lim_{L \rightarrow \infty} \frac{1}{L} \sum_{n=1}^{L}
 \mbox{ln} \mid \frac{\psi_{n+1}}{\psi_{n}} \mid  =  \lim_{L \rightarrow
\infty} \frac{1}{L} \sum_{n=1}^{L}
\mbox{ln}  \frac{r_{n+1}}{r_{n}}  \nonumber \\
 =  \frac{1}{2}\left< \mbox{ln} \left( 1 + A_n \sin2\theta_n + A_n^2\sin^2\theta_n \right)
\right>_n \; .
\label{eq:5a}
\end{eqnarray}
Here the brackets $\left<...\right>_n$ stand for averaging over $n$, i.e. along
the  trajectory of the map (\ref{eq:3}), (\ref{eq:4}). It is important to note
that the above expression for $\Lambda$ depends only on the angle $\theta_{n}$
and not on the radius $r_{n}$.  The above relation is correct for the energies
$E$ not very close to the
band edges $E=\pm 2$ where $\mu = 0,\pi$. At the band edges, there is additional
contribution to
the localization length that depends on the ratio $\sin \theta_{n+1}/\sin \theta_{n}$
\cite{IRT98}. In this paper we consider
the case when the energy $E$ is inside the allowed band, $|E| < 2$. Moreover, we
also exclude the band center $E=0$ where the localization length has a singular
behavior that requires
specific treatment (see \cite{IRT98} and references therein).

Apart from the abovementioned restrictions, the relation (\ref{eq:5a})
is valid for {\it any} potential $\epsilon_n$. However, analytical treatment
of the localization length is possible in the two limit cases of a weak or
strong potential.
Since in the following we consider the case of  weak disorder only,
$\mid A_n \mid \ll 1$, let us demonstrate
how the localization length can be derived from (\ref{eq:5a}).
We specify that the
distribution function of site energies $\epsilon_n$ is flat,
$P\left(\epsilon_{n}\right)=1/W$, within
the region $\mid \epsilon_{n} \mid \leq W/2$, with the variance
$\left<\epsilon_{n}^2\right>=W^2/12$.
Then, for $W \ll 1 $ the logarithm in (\ref{eq:5a}) can be expanded, and
by taking into account that in the lowest approximation with disorder
the values of $A_n$ and $\theta_n$ are
statistically independent, one can easily obtain,
\begin{equation}
l_{\infty}^{-1} = \frac{\left< \epsilon_{n}^{2} \right>}{8\sin^{2}\mu}
= \frac{W^{2}}{96\left( 1-\frac{E^2}{4}
\right)} \; .
\label{eq:5}
\end{equation}
For the first time this expression was derived by Thouless \cite{Thouless79}.
It works
quite well over the whole range of energies apart from the band center $E=0$.
The analytical treatment of the
expression (\ref{eq:5a}) was the main interest for different cases, including
correlated disorder. Unlike our previous
studies \cite{IKU01,IK99,KI00,Hmap,IRT98},
in what follows we address a question about the resistance of finite samples of size $L$.
Therefore, we are interested in the properties of the map (\ref{eq:2}) on the finite
time scale, for $n=1, ... , L$.

Our further consideration is based on the expression for the transmission coefficient
$T_L$ in terms of classical trajectories of the map (\ref{eq:2}) \cite{KTI97},
\begin{equation}
T_{L} = \frac{2}{1 + \frac{1}{2} \left( r_{1,L}^{2} + r_{2,L}^{2} \right)} \; \; .
\label{eq:6}
\end{equation}
Here $r_{1,L}$ and $r_{2,L}$ stand for the radii of the two
complimentary trajectories obtained by iterating the map
(\ref{eq:3}) up to the last site of the sample, $n=L$. Each of the
trajectories is specified by the initial conditions at $L=0$,
namely, $ r_{1,0} =  r_{2,0} = 1$, $\theta_{1,0}= 0$, and $
\theta_{2,0} = \pi/2$.

As one can see, all statistical properties of the transmission are entirely
determined by
the evolution of the two complementary trajectories of the classical map.
Some analytical and numerical analysis of the expression (\ref{eq:6})
have been recently performed in Ref. \cite{doss}. The question under study was
the statistical distribution of the transmission coefficient $T_L$.
In particular, it was shown that the
correlations between the two classical trajectories are different for
the ballistic and localized regimes, giving rise to the different
distribution functions of $T_L$. In next sections we address
the question of global
properties of the resistance $R_L=T_L^{-1}$, by paying main attention
to the mean
values $\left<R_L\right>$ and $\left<R_L^2\right>$.

\section{Resistance.}

\subsection{First moment.}

According to Eq. (\ref{eq:6}) the resistance $R_L=T_L^{-1}$ of a
sample of length $L$ is given by the following formula,
\begin{equation}
R_{L} = \frac{1}{2} + \frac{1}{4}r_{1, L}^{2} + \frac{1}{4}r_{2, L}^{2} .
\label{eq:Re1}
\end{equation}
Here the final radii $r_{i, L}^{2}$ ($i=1,2$) are expressed through the kicks
amplitudes $A_n$
and the phases $\theta_{n}^{(i)}\equiv \theta_{i,n}$ of
the two trajectories. Using Eq. (\ref{eq:4}) we get,
\begin{equation}
r_{i, L}^{2} = \prod_{n=0}^{L-1}\left(1 + A_{n}^{2}\sin^{2}\theta_{n}^{(i)} +
A_{n}\sin2\theta_{n}^{(i)} \right) \; .
\label{eq:rad1}
\end{equation}
It is convenient to represent the mean value of $r_{i, L}^{2}$ in the equivalent form
\begin{equation}
\left<r_{i, L}^{2}\right> = \left<\exp\left\{\sum_{n=0}^{L-1}\mbox{ln}\left(
1 + A_{n}^{2}\sin^{2}\theta_{n}^{(i)} + A_{n}\sin2\theta_{n}^{(i)} \right)
\right\}\right> .
\label{eq:mrad1}
\end{equation}
Then, in the limit of weak disorder, $\mid A_n\mid \ll 1$, one can
expand the logarithm in Eq. (\ref{eq:mrad1}). Separating quadratic and linear terms, we represent
$\left <r_{i, L}^{2}\right >$ in a form of the product of two factors,
\begin{eqnarray}
\left<r_{i, L}^{2}\right>  & = &
\left<\exp\left\{\sum_{n=0}^{L-1}A_{n}^{2}\sin^{2}\theta_{n}^{(i)}
- \frac{1}{2} \sum_{n=0}^{L-1}A_{n}^{2}\sin^{2}2\theta_{n}^{(i)}\right\}\right>
\nonumber \\
& & \times \left<\exp\left\{\sum_{n=0}^{L-1}A_{n}\sin2\theta_{n}^{(i)}\right\}\right> .
\label{eq:mrad2}
\end{eqnarray}
The first factor contains the sums
$\sum_{n=0}^{L-1}A_{n}^{2}\sin^{2}\phi_{n}^{(i)}$ where $\phi_{n}^{(i)}$ stands for
either $\theta_{n}^{(i)}$ or $2\theta_{n}^{(i)}$. Since $L >> 1$ these sums are self-averaged
quantities, with small variance. Thus, they can be substituted by their mean
values,
\begin{equation}
\sum_{n=0}^{L-1}A_{n}^{2}\sin^{2}\phi_{n}^{(n)}
\approx L\left<A_{n}^{2}\right>\left<\sin^{2}\phi_{n}^{(n)}\right>
= 4\lambda \; ,
\label{eq:msum1}
\end{equation}
with
\begin{equation}
\label{lam}
\lambda = \frac {L \left<A_{n}^{2}\right>}{8} = \frac {L}{l_{\infty}}.
\end{equation}
Therefore, the first factor of Eq.(\ref{eq:mrad2}) takes the form,
\begin{equation}
\left<\exp\left\{\sum_{n=0}^{L-1}A_{n}^{2}\sin^{2}\theta_{n}^{(i)}
+ \sum_{n=0}^{L-1}A_{n}^{2}\sin^{2}2\theta_{n}^{(i)}\right\}\right>
\approx \exp\left(2\lambda\right) \; .
\label{eq:1stfact}
\end{equation}

\begin{figure}[!tb]
\begin{center}
\leavevmode
\epsfysize=7cm
\epsfbox{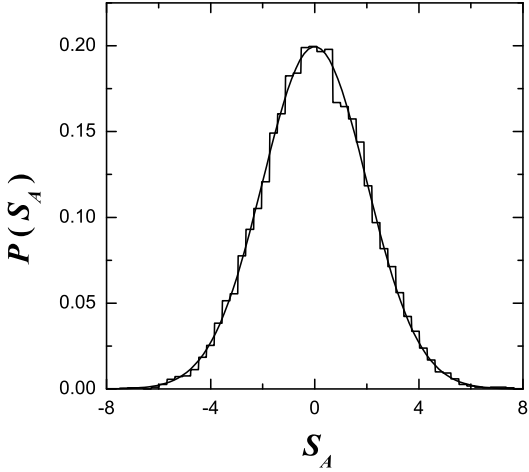}
\narrowtext
\caption{\small Histogram for the probability distribution of
$S_A=\sum_{n=0}^{L-1}A_{n}\sin2\theta_{n}^{(i)}$ plotted for $E=1.5$, $W=0.1$,
$L=4200$ and $\lambda=L/l_{\infty}=1$ for $10,000$ realizations of the random
potential $\epsilon_n$. The smooth curve is a Gaussian distribution
with variance $\sigma^{2}=4$.}
\label{fig:pasdh}
\end{center}
\end{figure}

The second factor in Eq.(\ref{eq:mrad2}) exhibits strong fluctuations.
We can calculate its mean value using the
distribution function ${\cal P(S_A)}$,
\begin{equation}
\left<f \right> = \int_{-\infty}^{\infty}f\left(S_A\right)
\, {\cal P}\left(S_A\right) \, dS_A \; ,
\label{eq:mvdef}
\end{equation}
where $f(S_A)=\exp \left(S_A \right)$ with
$S_A=\sum_{n=0}^{L-1}A_{n}\sin2\theta_{n}^{(i)}$. The variable
$S_A$ is a sum of random independent numbers, therefore, according
to the central limit theorem $P(S_A)$ is a Gaussian function with
variance $4\lambda$. The distribution of $S_A$ is shown in Fig.
\ref{fig:pasdh} and one can see quite good fitting of the
numerical data by a Gaussian function. As a result, the second
factor of Eq.(\ref{eq:mrad2}) can be calculated explicitly,
\begin{eqnarray}
\left<\exp\left(S_A\right)\right> & = & 
\frac{1}{\sqrt{8\pi\lambda}}
\int_{-\infty}^{\infty} \exp\left(S_A - 8\lambda^{-1}S_A^{2}\right) \, dS_A
\nonumber \\
= \exp\left(2\lambda\right).
\label{eq:2ndfact}
\end{eqnarray}

Substituting Eqs.(\ref{eq:1stfact}) and (\ref{eq:2ndfact}) into Eq.(\ref{eq:mrad2}),
we obtain the following expression for the mean value of $r_{i, L}^{2}$,
\begin{equation}
\left<r_{i, L}^{2}\right> = \exp\left(4\lambda\right) \; .
\label{eq:mrad3}
\end{equation}
This leads to the final formula for the mean value of the resistance
(\ref{eq:Re1}),
\begin{equation}
\left<R_L\right> = \frac{1}{2}\left[1 + \exp\left(4\lambda\right)\right] \; .
\label{eq:Re2}
\end{equation}
Note that this expression is valid for {\it any} value of the control parameter $\lambda$.
It gives the resistance in the metallic $(\lambda \ll 1)$ and insulator
regimes $(\lambda \gg 1 )$ as well as the crossover $(\lambda \simeq 1 )$.
The only condition that we have used here is the condition of weak disorder.

In the metallic regime, the fluctuations of the transmission coefficient
are weak and it is a self-averaged function. Therefore, the mean value
of conductance $\left< G \right >$ is the same as $1/ \left< R \right >$.
However, in the localized regime, both the conductance and resistance are
not self-averaged functions that lead to distinct values of  $\left< G \right >$
and $1/\left<R\right >$. Indeed, it is well-known that the mean conductance
has a pre-exponential factor $\sim (L/l_{\infty})^{3/2}$. Derivation of
this factor requires some special efforts \cite{LGP88}. This can be understood
from Eq. (\ref{eq:6}) which contains the random radii in the denominator.
Unlike this, in the case of resistance (\ref{eq:Re1}) these random variables
reside in the nominator and strongly simplify the calculations.

\subsection{Second moment.}

From Eq.(\ref{eq:Re1}) the second moment of the resistance can be expressed as follows,
\begin{equation}
R_{L}^2 = \frac{1}{4} + \frac{1}{4}r_{1, L}^{2} + \frac{1}{4}r_{2, L}^{2}
+ \frac{1}{16}r_{1, L}^{4} + \frac{1}{16}r_{2, L}^{4}
+ \frac{1}{8}r_{1, L}^{2}r_{2, L}^{2} \; .
\label{eq:2ndmomre}
\end{equation}
One can see that apart from the second and fourth moments of
$r_{i, L}$, this expression contains a product $r_{1, L}^{2}r_{2, L}^{2}$.
The mean value of this term depends on the correlations between the two
complementary trajectories of the classical map (\ref{eq:3}). This fact
strongly complicates analytical treatment. In what follows, we consider
the two limiting cases of the ballistic and localized regimes separately.

First, let us start with the evaluation of the mean value of the fourth
moments $r_{i, L}^{4}$. This
can be done by using Eq.(\ref{eq:rad1}) in the same way as described above,
directly and without restrictions,
\begin{eqnarray}
\left<r_{i, L}^{4}\right> & = &
\left<\exp\left(2\sum_{n=0}^{L-1}A_{n}^{2}\sin^{2}\theta_{n}^{(i)}
- \sum_{n=0}^{L-1}A_{n}^{2}\sin^{2}2\theta_{n}^{(i)}
\right)\right>
\nonumber \\
& & \times
\left<\exp\left(2\sum_{n=0}^{L-1}A_{n}\sin2\theta_{n}^{(i)}
\right)\right> \; .
\label{eq:mrads1}
\end{eqnarray}
Here we have used the weak disorder condition $A_n^2 \ll 1$ and have kept
only terms up to ${\cal O}\left(A_{n}^{2}\right)$. We have also presented the mean value in
the form of the product of two
factors, one with a small variance (first factor) and another that reveals
strong fluctuations (second factor). Inside the first factor, the
sum $\sum_{n=0}^{L-1}A_{n}^{2}\sin^{2}\phi_{n}^{(i)}$ with $\phi_{n}^{(i)}$ as
$\theta_{n}^{(i)}$ or $2\theta_{n}^{(i)}$, is a self-averaged
quantity. Therefore, after substitution by its mean value we get,
\begin{equation}
\left<\exp\left(2\sum_{n=0}^{L-1}A_{n}^{2}\sin^{2}\theta_{n}^{(i)}
- \sum_{n=0}^{L-1}A_{n}^{2}\sin^{2}2\theta_{n}^{(i)}\right)\right>
\approx \exp\left(4\lambda\right) \; .
\label{eq:1stfact2}
\end{equation}
For the second factor of Eq. (\ref{eq:mrads1}) we proceed in the same way as we
did in the derivation of Eq. (\ref{eq:2ndfact}). We use the fact that the quantity
$S_A = \sum_{n=0}^{L-1}A_{n}\sin2\theta_{n}^{(i)}$ is distributed according to the Gaussian
(see demonstration in Fig.\ref{fig:pasdh}) with the variance $4\lambda$.
Therefore, the mean value of the
second factor in Eq. (\ref{eq:mrads1}) is given by,
\begin{eqnarray}
\left<\exp\left(2S_A\right)\right>
& = & \frac{1}{\sqrt{8\pi\lambda}}\int_{-\infty}^{\infty}
\exp\left(2S_A - 8\lambda^{-1}S_A^{2}\right)\, dS_A 
\nonumber \\
& = & \exp\left(8\lambda\right) \; .
\label{eq:2ndfact2}
\end{eqnarray}
Substituting Eqs. (\ref{eq:1stfact2}) and (\ref{eq:2ndfact2}) into Eq. (\ref{eq:mrads1})
we obtain
the following result for the mean value of $r_{i, L}^{4}$,
\begin{equation}
\left<r_{i, L}^{4}\right> = \exp\left(12\lambda\right) \; .
\label{eq:mrads2}
\end{equation}

Let us now consider the term $r_{1, L}^{2}r_{2, L}^{2}$. We can write the
mean value of this
quantity in the following exact form,
\begin{eqnarray}
\nonumber 
\left<r_{1, L}^{2}r_{2, L}^{2}\right> =  \hspace{5cm} \\
\nonumber \\
\left<\exp\left\{\sum_{n=0}^{L-1}\mbox{ln}
\left( 1 + A_{n}^{2}\sin^{2}\theta_{n}^{(1)} + A_{n}\sin 2\theta_{n}^{(1)}
\right)\right\}\right.
\nonumber \\
& & \hspace{-7.5cm}
\times  \left. \exp\left\{\sum_{n=0}^{L-1}\mbox{ln}
\left( 1 + A_{n}^{2}\sin^{2}\theta_{n}^{(2)} + A_{n}\sin2\theta_{n}^{(2)}
\right)\right\}\right> \; .
\label{eq:r1r2a}
\end{eqnarray}
Using the weak disorder condition we can expand the logarithms and
present the correlator $\left <r_{1, L}^{2}r_{2, L}^{2}\right>$ as
follows,
\begin{eqnarray}
\left<r_{1, L}^{2}r_{2, L}^{2}\right> & = &
\left<\exp\left\{\sum_{n=0}^{L-1}A_{n}^{2} Z_n \right\} \right> \nonumber \\
& & \hspace{-2cm} \times \left<\exp\left\{\sum_{n=0}^{L-1}A_{n}\left(\sin2\theta_{n}^{(1)}
+ \sin2\theta_{n}^{(2)}\right)\right\}\right> 
\label{eq:r1r2b1} \\
\hspace*{-3cm} & \approx & 
\exp\left\{L\left< A_{n}^{2} \right> \left<Z_n\right> \right\}
\times \left<\exp\left(S_{P}\right)\right> \, ,
\label{eq:r1r2b}
\end{eqnarray}
where
\begin{equation}
Z_n = \left(\sin^{2}\theta_{n}^{(1)}
+ \sin^{2}\theta_{n}^{(2)}\right)  
- \frac{1}{2}\left(\sin^{2}2\theta_{n}^{(1)}
+ \sin^{2}2\theta_{n}^{(2)}\right)
\end{equation}
Here we introduced a random variable $S_{P}$,
\begin{equation}
S_{P}=\sum_{n=0}^{L-1}A_{n}\left(\sin2\theta_{n}^{(1)} + \sin2\theta_{n}^{(2)}\right),
\label{eq:S_P}
\end{equation}
and splited the mean value $\left<r_{1, L}^{2}r_{2, L}^{2}\right>$
into two factors, one with small variance and another with strong
fluctuations respectively, see Eq.(\ref{eq:r1r2b1}). Now, we have
to take into account the correlations between the phases
$\theta_{n}^{(1)}$ and $\theta_{n}^{(2)}$. In the first factor we
have the following self-averaged quantities:
$\sum_{n=0}^{L-1}A_{n}^{2}$,
$\sum_{n=0}^{L-1}\left(\sin^{2}\theta_{n}^{(1)} +
\sin^{2}\theta_{n}^{(2)}\right)$, and
$\sum_{n=0}^{L-1}\left(\sin^{2}2\theta_{n}^{(1)} +
\sin^{2}2\theta_{n}^{(2)}\right)$. The mean value of
$\left<A_{n}^{2}\right>$ is $8/l_{\infty}$, while the mean values
$\left<\sin^{2}\theta_{n}^{(1)} + \sin^{2}\theta_{n}^{(2)}\right>$
and $\left<\sin^{2}2\theta_{n}^{(1)} +
\sin^{2}2\theta_{n}^{(2)}\right>$ are very close to $1$.
Substituting these mean values in the first factor of
\mbox{Eq.(\ref{eq:r1r2b})} we get,
\begin{equation}
\left<r_{1, L}^{2}r_{2, L}^{2}\right>  =  \exp\left(4\lambda\right)
\left<\exp\left(S_{P}\right)\right> \; .
\label{eq:r1r2c}
\end{equation}

Now, we have to take into account the correlations between the phases
$\theta_{n}^{(1)}$ and $\theta_{n}^{(2)}$. These correlations is
the main problem for further analytical treatment. It turns out that
the correlations between phases $\theta_{n}^{(1)}$ and $\theta_{n}^{(2)}$
are very different in the ballistic ($\lambda \ll 1$) and localized ($\lambda \gg 1$)
regimes.

\subsubsection{Ballistic regime.}

In the ballistic regime the exponential $\exp{S_{P}}$ is close to one,
and we expand it up to quadratic terms,
\begin{eqnarray}
\nonumber \left<\exp\left(S_{P}\right)\right> \approx 1 + S_2 
\hspace{5cm} \\
+ \frac{1}{2}\left[\left<\sum_{n=0}^{L-1}A_{n}^{2}\sin^{2}2\theta_{n}^{(1)}\right> +
\left<\sum_{n=0}^{L-1}A_{n}^{2}\sin^{2}2\theta_{n}^{(2)}\right> 
\right]
\label{eq:2ndfact3}
\end{eqnarray}
Here the term $S_{2}$ has the following form,
\begin{equation}
S_{2} = 
\left< \sum_{n=0}^{L-1} A_{n}^{2} \sin \left( 2\theta_{n}^{(1)} \right)
\sin \left( 2\theta_{n}^{(2)} \right) \right> \; .
\label{eq:29x2}
\end{equation}
This term describes the correlations between the phases $\theta_{n}^{(1)}$ and
$\theta_{n}^{(2)}$ of the two classical trajectories that start from complementary
initial conditions.
It should be stressed that
the correlation term
keeps track of the change in the correlations along a finite sample, when running
from $n=0$ to $n=L-1$.
Taking into account that for the white-noise sequence $\epsilon_n$
the fluctuations of $A_{n}$ and $\theta_{n}$
are statistically independent \cite{IK99}, we can express $S_2$ as follows
\begin{equation}
S_{2}=8l_{\infty}^{-1}\sum_{n=0}^{L-1}R_{2}(\lambda_{n}) \; .
\label{eq:S2pc}
\end{equation}
Here we introduced a binary correlator $R_{2}$
\begin{equation}
R_{2}=\left< \sin \left( 2\theta_{n}^{(1)} \right)
\sin \left( 2\theta_{n}^{(2)} \right) \right>
\label{eq:2pc}
\end{equation}
that is expected to depend on the scaling parameter
$\lambda_{n}=n/l_{\infty}$ only.
Note that the average in Eq.(\ref{eq:S2pc}) is taken over different
realizations of the disordered
potential for a {\it fixed} value of $n=1,...,L-1$.
Due to the ergodicity, this average is performed over different
realizations of $\epsilon_n$
after running along the trajectory up to a fixed $n$.

\begin{figure}[!tb]
\begin{center}
\leavevmode
\epsfysize=7.15cm
\epsfbox{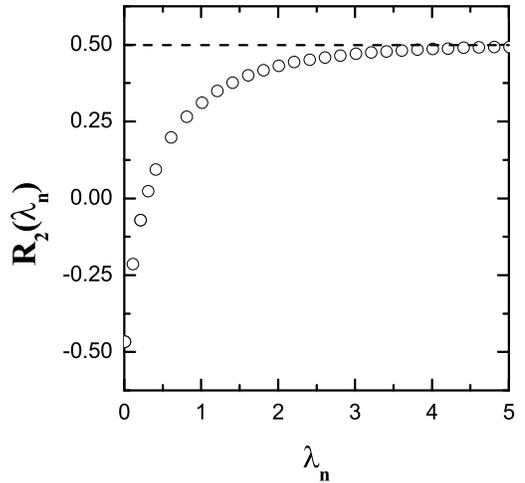}
\narrowtext
\caption{Numerical data for the correlator $R_2$ versus
scaling parameter $\lambda_n = n/l_\infty $ is shown for the transition
from the ballistic to localized regime for $E=1.5$
and $W=0.1$. The average was done over $10^4$ realizations of the
disorder with an additional "window moving" average, in order
to reduce large fluctuations.}
\label{fig:2}
\end{center}
\end{figure}

A specific character of the correlator $R_2$ is clearly seen from the
data reported in Fig. \ref{fig:2}.
As one can see,
the correlator $R_{2}$ changes from $-1/2$ in the ballistic regime,
$\lambda_{n} \ll 1$, to $1/2$ in the localized regime, $\lambda_{n} \gg 1$.
Taking into account that $\left<\sin^2 \theta_n \right > =1/2$,
the latter means synchronization of the phases in the localized regime,
$\theta_{n}^{(1)} \approx \theta_{n}^{(2)}$, see details
in \cite{doss}. In the ballistic regime the phases are anti-synchronized,
$\theta_{n}^{(1)} \approx - \theta_{n}^{(2)}$, however, this holds only
for very small values of $\lambda$ as one can see from the graph in
Fig. \ref{fig:2}.  This behavior is a manifestation of peculiar
statistical correlations between the
phases $\theta_{n}^{(1)}$  and $\theta_{n}^{(2}$.

In order to evaluate analytically the correlator $R_{2}$ in the ballistic regime,
we use the approximate map for the phase $\theta_{n}$ that can be directly
obtained from
\mbox{Eq.\ (\ref{eq:3})} in the limit of weak disorder,
\begin{equation}
\theta_{n+1} = \theta_{n} - \mu + \epsilon_{n} \frac{\sin^2 \theta_{n}}{\sin \mu} \; .
\label {eq:11}
\end{equation}
Iterating this map, one can express an angle $\theta_{n}$ in terms
of the amplitudes $\epsilon_{0},\epsilon_{1},...,\epsilon_{n-1}$ of all
previous kicks for a fixed
value of $\mu$ (energy $E$). Taking into account the initial conditions
\mbox{$\theta_{0}^{(1)}  = 0$} and
\mbox{$\theta_{0}^{(2)} = \pi/2 $}, the following explicit formulas are derived
for the phases,
\begin{eqnarray}
\begin{array}{l}
\theta_{n}^{(1)} \approx -n\mu + \sin^{-1}\mu \sum_{i=0}^{n-1}
\epsilon_{i}\sin^{2}(i \mu) 
\vspace{0.3cm} \\
- \sin^{-2}\mu \sum_{i=0}^{n-1} \epsilon_{i}\sin^3(i\mu)\cos(i\mu) 
\vspace{0.3cm} \\
- \sin^{-2}\mu \sum_{i<k}^{n-1}
\epsilon_{i} \epsilon_{k} \sin^2(i\mu)\sin(2k\mu) + {\cal O}
\left( \epsilon^{3} \right) ,
\end{array}
\label{eq:11a}
\end{eqnarray}
and
\begin{eqnarray}
\begin{array}{l}
\theta_{n}^{(2)} \approx \frac{\pi}{2} -n\mu + \sin^{-1}\mu \sum_{i=0}^{n-1}
\epsilon_{i}\cos^{2}(i \mu) 
\vspace{0.3cm} \\
- \sin^{-2}\mu \sum_{i=0}^{n-1} \epsilon_{i}\cos^3(i\mu)\sin(i\mu) 
\vspace{0.3cm} \\
- \sin^{-2}\mu \sum_{i<k}^{n-1}
\epsilon_{i} \epsilon_{k} \cos^2(i\mu)\sin(2k\mu) + {\cal O}
\left( \epsilon^{3} \right)
\end{array}
\label{eq:11b}
\end{eqnarray}

We substitute the above expansions  into Eq. (\ref{eq:2pc}) for the correlator $R_{2}$.
Neglecting the contributions of the fast oscillating terms
$ \sin^{m} \left( k n \mu \right) $ with positive integers
$k$ and $m$, after some algebra we get,
\begin{equation}
R_{2} (\lambda_{n}) = - \frac{1}{2} + 4 \lambda_{n} - 16 \lambda_{n}^{2}
+ {\cal O}\left(\lambda^{3}\right) \; .
\label{eq:12}
\end{equation}
Numerical data in Fig. (\ref{fig:2}) confirm the validity of this estimate
in the ballistic regime,
$\lambda_{n} \ll 1$ (see also \cite{doss}).
Then the two-point correlator $S_{2}$ takes the form,
\begin{equation}
S_{2}
= - 4 \lambda + 16 \lambda^{2} + {\cal O} \left(\lambda^{3}\right) \; ,
\label{eq:12sum}
\end{equation}

Now we come back to Eqs. (\ref{eq:r1r2c}), (\ref{eq:2ndfact3}), and (\ref{eq:29x2}).
The mean values of the two sums in Eq. (\ref{eq:2ndfact3}) can be easily evaluated
and for the correlator
$\left <r_{1, L}^{2}r_{2, L}^{2}\right>$ we obtain,
\begin{equation}
\left<r_{1, L}^{2}r_{2, L}^{2}\right>  =  \left[1+16\lambda^{2}
+ O\left(\lambda^{3}\right) \right]\exp\left(4\lambda\right)
 \; .
\label{eq:r1r2d}
\end{equation}

\begin{figure}[!htb]
\begin{center}
\leavevmode
\epsfysize=7cm
\epsfbox{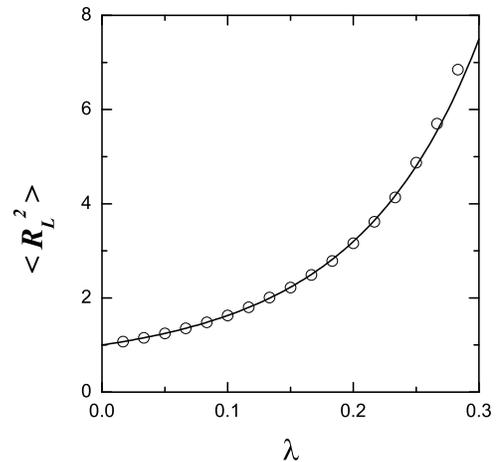}
\narrowtext
\caption{\small Plot of Eq.\ (\ref{approx}) (solid line) against numerical data
(open circles) within the ballistic regime. Numerical data are
obtained for $E=1.5, W=0.1, l_{\infty}=4200$ and different lengths $L$,
with the average over $50,000$ realizations of disorder. }
\label{fig:Re2MR}
\end{center}
\end{figure}

Now the second moment of the resistance in the ballistic regime can be expressed as,
\begin{equation}
\left<R_{L}^2\right> = \frac{1}{8}\left[2 + \left(5+16\lambda^{2}\right)
\exp\left(4\lambda\right) + \exp\left(12\lambda\right)\right] \; .
\label{eq:Re2MR}
\end{equation}
Since this expression is valid in the quadratic approximation with $\lambda^2$,
we expand the exponential terms and obtain the final formula for the second
moment of resistance in the ballistic regime,
\begin{equation}
\left<R_{L}^2\right> = 1 + 4 \lambda + 16 \lambda^2 + O\left(\lambda^{3}\right) \; .
\label{approx}
\end{equation}
There is a good agreement between this analytical result and numerical curve shown in
Fig. \ref{fig:Re2MR} up to $\lambda \approx 0.3$.

\subsubsection{Strongly localized regime.}

\begin{figure}[!htb]
\begin{center}
\leavevmode \epsfysize=7cm \epsfbox{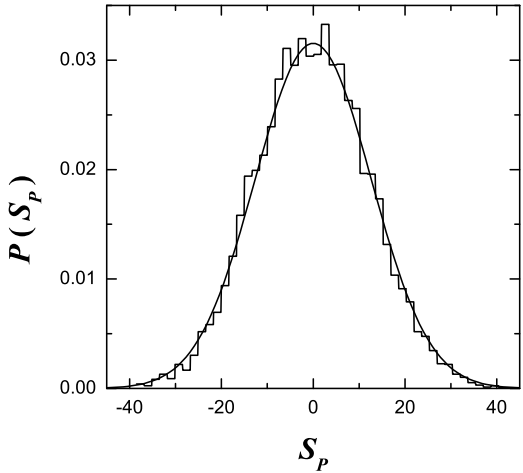}
\narrowtext
\caption{\small
Numerical data (solid line) for the distribution of $S_P$ plotted
against the Gaussian (dotted line) with the zero mean and variance
$\sigma^{2}=16$. The histogram for the distribution of $S_P$ fits
well by a Gaussian curve. Numerical data are shown for $E=1.5,
W=0.1, L=42000$ and $L/l_{\infty}=10$, with additional average
over $10^4$ realizations of the random potential.}
\label{fig:pas2ts2pdh}
\end{center}
\end{figure}

Let us now consider strongly localized regime, $\lambda \gg 1 $.
As was mentioned, in this regime the phases  fluctuate coherently,
i.e. $\theta_{n}^{(1)} \approx \theta_{n}^{(2)}$.
This property helps to evaluate the mean value $\left<S_P\right>$, see
Eq. (\ref{eq:S_P}). Using the same procedure that was used for
the ballistic regime, we find
that in the localized regime the distribution function of random
variable $S_{P}$ is the Gaussian with zero mean value and variance
\begin{equation}
\left<S_{P}^{2}\right> - \left<S_{P}\right>^{2} = 16\lambda \; .
\label{eq:varsum}
\end{equation}
Numerical data in Fig. \ref{fig:pas2ts2pdh} confirm this analytical result.

The mean value of the normally distributed random variable $S_P$ can be easily evaluated,
\begin{equation}
\left<\exp\left(S_{P}\right)\right> = \exp\left(8\lambda\right) \; .
\label{eq:2ndfact5}
\end{equation}
Then, substituting Eq.(\ref{eq:2ndfact5}) into Eq.(\ref{eq:r1r2c}),
we get the analytical expression for the
mean value of the correlator $r_{1, L}^{2}r_{2, L}^{2}$,
\begin{equation}
\left<r_{1, L}^{2}r_{2, L}^{2}\right> = \exp\left(12\lambda\right) \; .
\label{eq:r1r2e}
\end{equation}
In the localized regime this correlator is exponentially large.
This means that the trajectory of the map Eq. (\ref{eq:2})
extends far away from the origin \cite{Hmap}. Now it is easy to
show that the variance $R_L^2$ also grows exponentially in the localized regime,
\begin{equation}
\left<R_L^2\right> = \frac{1}{4} \exp \left(12\lambda \right)
+\frac{1}{2} \exp\left(4\lambda\right)
+ \frac{1}{4} \; .
\label{eq:Re2SLR}
\end{equation}

Direct numerical computation of the mean value $ \left<R_L^2\right> $
gives rise to serious
problems, since in a strongly localized regime the variance as well
as all the moments of $R_L^2$
fluctuate enormously with realization of the disordered potential.
This is a reflection
of the well known fact that $R_L$ (as well
as the conductance $T_L$) is not a self-averaged quantity. Instead,
it is more physical
to make an average of the logarithm of the moments.

\section{Discussion and concluding remarks}

To the best of our knowledge, there is no systematic study of the statistical properties of
the resistance $R_L$ for the tight-binding models. Only brief discussion of the
moments of $R_L$ can be found in Ref. \cite{Makarov}, where the analysis was done
with the use of the
''two-scale approach".

In the spirit of the discussion of the single parameter scaling \cite{alt}
of the conductance,
it is also interesting to consider the first and second moments of $\ln(R_S)$.
Due to the simple
relation $R_L=T_L^{-1}$ between the resistance and conductance, it is clear
that in the strongly
localized regime, where the SPS is assumed to work well, the results \cite{doss}
obtained for $T_L$
are also valid for $R_L$. Indeed, since $\ln(R_L) = - \ln(T_L)$, the following
relations hold
\begin{equation}
\label{loglog}
\left <\ln R_L \right > = 2\lambda; \,\,\,\,\,\,\,\,\,\,
\left < \ln ^2 R_L \right> = 4\lambda + 4\lambda^2 \; ,
\end{equation}
\begin{equation}
Var \left( \ln R_L \right) = \left < \ln ^2 R_L \right>
- \left <\ln R_L \right > ^2 =
2 \left <\ln R_L \right > \; .
\label{spsR}
\end{equation}
The last formula corresponds to the single parameter scaling for resistance.
We remind that the SPS is applied for conductance $T_L$ \cite{alt} and
results in the relation,
\begin{equation}
{Var} \left( \ln T_L \right) = - 2 \left <\ln T_L \right > \; .
\label{spsT}
\end{equation}

In conclusion, in this paper we have analytically studied the main
properties of the resistance $R_L$ of finite 1D samples with
random potentials. Our analysis is based on the Hamiltonian
version of the transfer matrix method. The main equation of this
method is the Hamiltonian map that replaces the discrete
Schrodinger equation. In this way all statistical properties of
resistance can be expressed through two complimentary trajectories
of this map. Statistical correlations between these trajectories
play an essential role and an analytical evaluation of the moments
of the resistance is possible in the situations when the binary
correlator of phases is known. At the same time, the mean value of
$R_L$ is independent of this correlator. Therefore, the expression
for $\left< R_L \right> $ can be found exactly, in the whole range
of the scaling parameter, $0< \lambda = L/l_\infty < \infty$. In
contrast, the mean value of the second moment, $\left< R_L^2
\right> $, depends on the binary correlator $R_2$. Analytical
evaluation of this term was found to be possible in the limiting
two cases of ballistic and localized regimes. The derived
analytical expressions for the resistance and its second moment
correspond quite well to the numerical data obtained in a wide
range of the model parameters.

\section{Acknowledgments}

The authors are very thankful to Luca Tessieri for his valuable
comments and fruitful discussions. This research was supported by
Consejo Nacional de Ciencia y Tecnolog\'{\i}a (CONACYT, M\'exico)
grants 34668-E and 42136-F.

\end{multicols}
\end{document}